\documentclass[aps,prl,twocolumn,showpacs,preprintnumbers,amsmath,superscriptaddress]{revtex4}
\usepackage{bm}
\usepackage{amsmath}
\usepackage[dvips]{graphicx}

\begin{document}
\bibliographystyle{apsrev}

\title{Theory  of incommensurate magnetic correlations across the
insulator-superconductor transition in  underdoped
La$_{2-x}$Sr$_x$CuO$_4$}
\author{Oleg P. Sushkov}
\email{sushkov@phys.unsw.edu.au}
\affiliation{School of
Physics, University of New South Wales, Sydney 2052, Australia}
\author{Valeri N. Kotov}
\email{valeri.kotov@epfl.ch}
\affiliation{Institute of Theoretical Physics, 
Swiss Federal Institute of Technology (EPFL), \\
CH-1015 Lausanne, Switzerland}

\date{\today}
\begin{abstract}
The main feature in the elastic neutron scattering of La$_{2-x}$Sr$_x$CuO$_4$ is the
existence of incommensurate peaks with positions that jump from 45$^\circ$ to 0$^\circ$ at 5\%
doping. We show that the spiral state of the $t-t'-t''-J$ model with realistic 
parameters describes this data  perfectly.
We explain why in the insulator the peak is at 45$^\circ$ 
while it switches to  0$^\circ$  precisely
at the insulator-metal transition.
The calculated positions of the peaks are in agreement with the data in both phases.
\end{abstract}
\maketitle

{\it Introduction.}
The phase diagram of La$_{2-x}$Sr$_x$CuO$_4$ (LSCO) shows that the magnetic state
changes dramatically with Sr doping. The parent compound 
La$_{2}$CuO$_4$ exhibits three-dimensional long-range antiferromagnetic (AF)
order below 325K \cite{KeimerNeel}. The N\'eel order disappears at  Sr
concentration $x\approx 0.02$,  however two-dimensional
(2D) short-range AF correlations exist at any doping \cite{Keimer} (see also
Ref.~\cite{Kastner} for a review).  At $x \leq 0.055$ the system exhibits
only hopping conductivity and behaves like an Anderson insulator, while the
usual dc conductivity as well as superconductivity appear at $x > 0.055$ \cite{Keimer,Kastner}.

Static magnetic ordering at very low temperatures has been observed both for
$x < 0.055$ and $x > 0.055$. The elastic neutron scattering peak is close
to the AF position ${\bf Q}_0=(\pi,\pi)$, but is shifted
from this position by $\delta{\bf Q}$:  
 ${\bf Q} = {\bf Q}_0 + \delta{\bf Q}$. We set the lattice spacing $a=1$. 
This shift indicates a one-dimensional incommensurate spin modulation.
The dependence of the shift on doping has been studied
in the superconducting phase \cite{Yamada}, as well as in 
the insulating phase \cite{Wakimoto,Matsuda,Fujita}.
These studies have revealed the following remarkably simple dependence of the elastic peak shift
on doping $x$ (see Fig.~7 in \cite{Fujita}):
\begin{eqnarray}
\label{data}
0.055 < x < 0.12:  && \delta{\bf Q} \approx 2 x  (\pm \pi,0) \ \mbox{or}  \ 
\delta{\bf Q} \approx 2 x (0,\pm \pi)   \ ,\nonumber\\
0.02 < x < 0.055:  && \delta{\bf Q} \approx \sqrt{2} x
(\pm\pi,\mp\pi)  \ .
\end{eqnarray}
Thus the 1D incommensurate spin modulation is proportional to doping and the direction
jumps from 45$^\circ$ to 0$^\circ$ exactly at the point of the insulator-metal transition.

One of the early proposals made by Shraiman and Siggia in Ref.~\cite{SS}, and later 
 explored in the context of the Hubbard and the t-J models
\cite{Kane,Auer,Zlatko,IF,Kampf,CM,SK},
 was that for small doping the collinear  N\'eel order gives way to a non-collinear spiral
 state. There is a gain in energy  since the holes can hop easier in a spiral background. 
 However the issue of stability of the spiral state remained rather controversial.
Using  chiral perturbation theory \cite{CP} we have recently revisited the problem of  
stability of the spiral state in the extended $t-t'-t''-J$ model \cite{SK},  
and have found that the uniform (1,0) spiral state is stable (at low doping) above some
critical values of $t',t''$. The stability is due to quantum fluctuations (order from disorder
effect). Even more importantly, superconductivity coexists with the spiral order.
The starting point of the approach \cite{SK} is the ground state of the Heisenberg model 
which incorporates all spin quantum fluctuations.
The chiral perturbation theory allows  a regular calculation of all physical
quantities in the leading  order approximation in powers of doping $x$. Subleading powers
of $x$ depend on the short-range dynamics and hence cannot be calculated without uncontrolled 
approximations. Therefore the approach is parametrically well justified  in the limit
 $x \ll 1$.
The phase diagram of the $t-t'-t''-J$ model obtained in Refs.~\cite{SK} is presented
in Fig.~1. 
 From the Raman 
data \cite{Tokura} $J \approx 125meV$ and we  set $t/J =3.1$,
following the calculations of Andersen {\it et al} \cite{And}.
The values  $t'\approx -0.5J$, $t''\approx 0.3J$ for LSCO and
 $t'\approx -0.8J$, $t''\approx 0.6J$ for
YBCO are taken from the same calculation.
From now on  we measure all energies in units of $J$ ($J=1$). 
 The matrix elements $t'$ and $t''$
are small compared to $t$, but nevertheless  are crucially important for
the  stability
because they  influence substantially the hole dispersion.
\begin{figure}
\centering
\includegraphics[height=140pt,keepaspectratio=true]{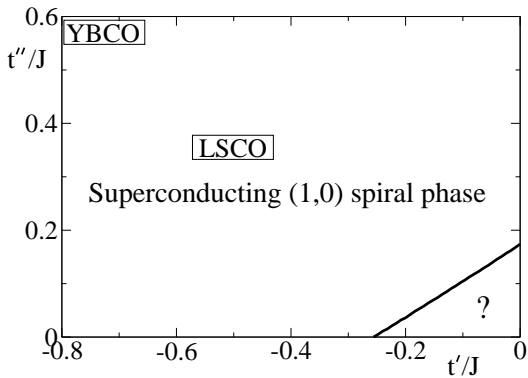}
\caption{The phase diagram of the $t-t'-t''-J$ model at $t/J=3.1$ and small 
{\bf uniform} doping \cite{SK}.
Points corresponding to LSCO and YBCO are shown. The line separates the stable
 spiral phase and  the unstable region (labeled ``?'').}
\label{Fig1}
\end{figure}
\noindent
The pitch of the uniform superconducting spiral state is \cite{SK}
\begin{equation}
\label{qc}
\delta {\bf Q}=\frac{Zt}{\rho_s}\ x \ (1,0) 
\approx 5.8 \ x \ (1,0) \ ,
\end{equation}
where $\rho_s\approx0.18$ is the spin stiffness of the Heisenberg model and 
$Z\approx0.34$ is the quasiparticle residue. 
The residue  depends weakly on $t'$ and $t''$, and  0.34 is the value for LSCO  \cite{SK}.
Notice that Eq.~(\ref{qc}) is in very good agreement with the  data (\ref{data}) in the
metallic phase, $x > 0.055$.
However different analysis is needed to  explain the data (\ref{data}) in the insulating
region,  $x < 0.055$.

In what follows we will  consider mostly the  insulating phase. 
The possibility of spiral ordering in the insulator has been stressed recently by
Hasselmann {\it et al} \cite{HCS}. However, the dynamical origin of the spiral,
the pitch of the spiral as well as the ``jump'' of the spiral direction at the insulator-metal transition 
still remain unexplained. It is the purpose of the present work to explain all these phenomena. 
In essence our idea is the following. The spiral (1,0)-state of the conductor
has lower energy  than that of the (1,1)-state only due to the presence of the Fermi motion
energy \cite{SK}.
On the other hand without the Fermi motion contribution the (1,1)-energy is lower than 
the (1,0)-energy. 
In the region $x < 0.055$ each hole is localized near its Sr ion, 
hence there is no Pauli blocking (Fermi energy) and  the system immediately  crosses over to
the (1,1) spiral state. Following this line of reasoning we will
demonstrate below that the whole variety
of experimental data in the insulator and across the insulator-metal
boundary can be consistently explained.
In the insulator the ground state is strongly non-uniform, with the  holes trapped 
 in the vicinity of the randomly distributed Sr ions, as confirmed by the
variable range  hopping (VRH) behavior of the dc
 conductivity for   $x < 0.055$ \cite{Kastner,LG}.
Due to the presence of the Coulomb potential (see below) and
 disorder, it seems likely that  the transition 
 to the metallic (uniform) phase at $x=0.055$ is of the density-driven percolation type.
 Thus we adopt this point of view  although the exact
 nature of the transition is not crucially important for our analysis.

{\it Coulomb trapping of holes.}
Let us consider first a single Sr ion with a single hole in an AF background.
The hole is trapped near Sr by the Coulomb potential $e^2/(\epsilon_{e}\sqrt{r^2+d^2})
\approx e^2/(\epsilon_{e}r)$ where $d$ is the distance from the CuO$_2$ plane to the Sr
ion and $\epsilon_{e}$ is the effective dielectric constant. For zero doping 
$\epsilon_e \sim 30$ and it increases  with doping, as discussed in 
\cite{Kastner}. In  momentum space the hole is localized near one of
the points  ${\bf k}_0=(\pm\pi/2,\pm\pi/2)$, which are the centers of the four faces 
of the magnetic Brillouin zone (MBZ). 
In the vicinity of these points the dispersion is quadratic:
$\epsilon_{\bf k}\approx \frac{\beta_1}{2}k_1^2+\frac{\beta_2}{2}k_2^2$, 
where ${\bf k}$ is defined with respect to ${\bf k}_0$, and $k_1$ is perpendicular
to the face of the MBZ, while $k_2$ is parallel to it. 
For values of $t'$ and $t''$
corresponding to LSCO we find that  the dispersion is practically isotropic 
$\beta_1\approx\beta_2=\beta\approx 2.2$ \cite{SK}.
Since the lattice spacing is about  $3.85\textrm{\AA}$ this value corresponds to an effective 
mass of about two free electron masses, in agreement with the  optical conductivity data 
\cite{Kastner}. The solution of the Schr\"odinger equation
\begin{equation}
\label{Sch}
\left(-\beta\nabla^2/2 -e^2/(\epsilon_er)\right)\chi=\epsilon\chi \ 
\end{equation}
determines the ground state wave function and the ground state energy of the localized hole:
\begin{eqnarray}
\label{wf}
\chi(r)= \sqrt{2/\pi}\kappa e^{-\kappa r} \ , \ \ \epsilon=-\beta\kappa^{2}/2 \,
\end{eqnarray}
where $\kappa=2e^2/(\epsilon_e\beta)$.
Using the hopping conductivity data  at
very low doping ($x=0.002$) \cite{Chen} we estimate the inverse size  $\kappa \approx 0.4$.
At higher doping (but still in the insulator) the value of $\kappa$ might decrease
slightly.
Thus throughout the insulating phase 
 $\kappa $ is small and this justifies the semiclassical
approximation we use below. Note that  the semiclassical approximation is used 
only with respect to the kinetic energy of the hole but not for the
spin (we do not use the 1/S expansion and account for all spin quantum fluctuations
via the chiral perturbation theory \cite{SK}).

{\it Spiral induced by a single trapped hole.}
Equations  (\ref{Sch}) and (\ref{wf}) assume a rigid antiferromagnetic background.
However one can gain  energy relaxing the background into the spiral state.
In the spiral state there are still two sublattices, 
sublattice ``up'' and sublattice ``down'', but the spin at every site  of each sublattice is 
rotated by an angle $\theta_i$ with respect to the orientation at $r=\infty$
\begin{eqnarray}
\label{spir}
|\mbox{i}\rangle&=&e^{i\theta({\bf{r_i}}){{\bf{m}}}
\cdot{\bf{\sigma}}/2}|\uparrow\rangle,   \
\mbox{ \  i $\in$  ``up'' \ sublattice}  ,\nonumber\\
|\mbox{j}\rangle&=&e^{i\theta({\bf{r_j}}){\bf{m}}
\cdot{\bf{\sigma}}/2}|\downarrow\rangle,   \
\mbox{ \  j $\in$  ``down'' \ sublattice} .
\end{eqnarray} 
Here ${\bf m}= (\cos\alpha,\sin\alpha,0)$ with arbitrary $\alpha$  is  the ``director'' of the spiral
which is orthogonal to the magnetization plane.
Note that directions in  spin space are completely independent of directions
in  coordinate space. 
The wave function of the hole $\psi({\bf r})$ has two components corresponding to
up and down sublattices. The total  energy is of the form \cite{IF,SK}
\begin{eqnarray}
\label{Hab1} 
&&E=\int d^2r \left\{\frac{\rho_s}{2} (\nabla\theta)^2+\right.\\
&&\left. \psi^{\dag}({\bf r})\left(\begin{matrix}
-\beta\frac{\nabla^2}{2}-\frac{e^2}{\epsilon_er} \\
\sqrt{2}Z t e^{i\alpha} ({\bf e}\cdot{\bf \nabla}\theta)
\end{matrix}
\begin{matrix}
\sqrt{2}Z t e^{-i\alpha} ({\bf e}\cdot\nabla\theta)\\
-\beta\frac{\nabla^2}{2}-\frac{e^2}{\epsilon_er}
\end{matrix}\right) \psi({\bf r}) \right\} \ ,\nonumber
\end{eqnarray}
where ${\bf e}=(\frac{1}{\sqrt{2}},\pm \frac{1}{\sqrt{2}})$
is a unit vector orthogonal to a given face of the MBZ.
Let us search for a solution in the form
\begin{equation}
\label{psis}
\psi({\bf r})=\frac{1}{\sqrt{2}}
\left(\begin{matrix}
1 \\
-e^{i\alpha}
\end{matrix}
\right)\chi(r)
\end{equation}
where $\chi(r)$ is given by Eq.~(\ref{wf}). Variation of the energy (\ref{Hab1})
with respect to $\theta$ leads to the following equation
\begin{equation}
\label{eqph}
\nabla^2 \theta =\sqrt{2}\frac{Zt}{\rho_s} ({\bf e}\cdot{\bf \nabla})\chi^2(r) \ .
\end{equation}
The solution of (\ref{eqph}) is
\begin{equation}
\label{solthet}
\theta=\frac{Zt}{\sqrt{2}\pi\rho_s}\frac{({\bf e\cdot r})}{r^2}
\left[1-e^{-2\kappa r}(1+2\kappa r)\right] \ .
\end{equation}
Substitution of this solution together with (\ref{psis}) and (\ref{wf}) in Eq.~(\ref{Hab1})
gives the following total energy
\begin{equation}
\label{ev1}
E=\left(\frac{\beta}{2}-\frac{Z^2t^2}{4\pi\rho_s}\right)\kappa^2
-2\frac{e^2}{\epsilon_e}\kappa \ .
\end{equation}
Minimizing this energy one finds $\kappa=\frac{2e^2}{\epsilon_e}/
[\beta-Z^2t^2/(2\pi\rho_s)]$.
However we  do not use directly this expression  since 
the effective dielectric constant $\epsilon_e$ is not known accurately enough.
Instead  we rely on estimates for $\kappa$ which directly follow from the hopping 
conductivity 
as  discussed after Eq.~(\ref{wf}).
As can be easily seen from Eqs.~(\ref{spir},\ref{solthet}), 
at distances $r \ll 1/\kappa$ our solution describes a $(1,\pm1)$ spiral,
 while in the opposite limit an effective dipole is formed (see discussion below).
The solution  is a variational one
because we have used the ansatz (\ref{wf}). 
Even though one can easily derive an exact  equation for $\chi$ which can be
solved numerically, this is not necessary since for our purposes the 
details of the charge distribution are not important.

We emphasize that the Coulomb trapping of the hole is crucially important. Without such trapping the hole
is delocalized and a single delocalized hole does not generate a static spiral.
This is qualitatively different from the arguments of Ref.~\cite{SS1}.
The solution (\ref{solthet}) does not carry any topological numbers, and consequently,  unlike  the model
used in Ref.~\cite{Gooding}, our solution  is {\it not} a skyrmion.
Other topological reasons for ``self-trapping'' of holes have also been
 given \cite{Kou}, however  we  pursue the Coulomb trapping picture since it unambiguously
follows from the parametrically justified analysis of the t-J model.


The solution (\ref{spir}),(\ref{solthet}) depends on the spiral ``director''
which is  
 a purely classical variable and
the energy is independent of it.
It is unlikely that  a finite system has an exactly  degenerate ground state
(spontaneous violation of symmetry).
This means that  higher orders in $\kappa$ ($\kappa^4$-corrections to the
semiclassical solution)  may give rise to a kinetic energy for the
director ${\bf m}$ and hence to quantum rotations of ${\bf m}$, lifting the 
  degeneracy. Another quantum effect is tunneling from one pocket
in momentum space to another.
However, the quantum corrections are not important 
for understanding the properties of LSCO since at finite  
concentration of impurities the interaction between them is much more important 
than the quantum  corrections to the semiclassical limit.

{\it  Effective dipole moment of the impurity
and destruction of the N\'eel order at 2\% doping.}
It is convenient to rewrite Eq.~(\ref{solthet}) using  the notation of the non-linear
$\sigma$-model. Far from the impurity core, $r \gg 1/\kappa$, the solution reads
\begin{equation}
\label{dn}
\delta {\bf n}=\overline{\bf m}\theta =\overline{\bf m} M
\frac{({\bf e\cdot r})}{2\pi r^2} \ \ , \ \ 
M=\frac{\sqrt{2}Zt}{\rho_s} \approx 8.2\
\end{equation}
where ${\bf n}=\delta {\bf n}+{\bf n_0}$ is the unit vector of antiferromagnetism,
${\bf n_0}={\bf n}(r=\infty)$, $\overline{\bf m}=[{\bf n_0}\times{\bf m}]$,
and ${\bf m}$ is the director of the impurity.
Here $M$ is the effective dipole moment of the impurity.
Note that  it is very large, $M \gg 1$.

The idea of destruction of the N\'eel order by randomly quenched dipoles was 
put forward by Glazman and Ioselevich \cite{Glazman}.
Detailed renormalization group calculations based on this picture have been performed 
by Cherepanov {\it et al} \cite{Cherepanov}
and we use their results. In particular an analysis of the experimental data by
Keimer {\it et al} \cite{Keimer} for the  in-plane correlation length at $x < 0.02$
was performed in \cite{Cherepanov}.
This analysis shows that in order to explain the data and hence the destruction of the
N\'eel order at $x\approx 0.02$ one needs to have a  value of $M$ which satisfies the
following condition \cite{Cherepanov}
\begin{equation}
\label{mexp}
A=\frac{M^2}{{\cal N}d}=20(1\pm 0.3) \ .
\end{equation}
Here $d=2$ is the  dimensionality of the problem and ${\cal N}$ is the dimensionality of the vector 
$\overline{\bf m}$. In our theory ${\cal N}=2$ because $\overline{\bf m}$ is orthogonal 
to ${\bf n}_0$. Hence we conclude from (\ref{mexp}) that $M_{exp}=8.9 (1\pm 0.15)$. 
This agrees  well with the theoretical value (\ref{dn}).

{\it Structure of the insulating (spin glass) region and transition into the metallic phase.}
Here we consider the range of doping $0.02 < x < 0.055$ where the insulating spin glass state
is realized.
Since elastic incommensurate neutron peaks have been observed in this regime \cite{Wakimoto,Matsuda,Fujita},
 there are two characteristic length scales:  $l_I\propto 1/x$,  related to the 
incommensurability, and the  magnetic correlation length $l_{M} > l_I$,  related to the spin 
glass disorder (randomness) and reflected in the (inverse) width of the elastic
 neutron peaks.

It is clear that  in order to  minimize the dipole-dipole interaction energy at finite impurity  
concentration (and at zero temperature), the dipoles (\ref{dn}) will align in such a way 
that all vectors ${\bf e}$ and ${\overline {\bf m}}$ are the same. 
 Such an  alignment is possible in spite 
of the  random positions and  generates an average spiral \cite{HCS}. Certainly around 
each  dipole there are deviations from the average described by (\ref{solthet}).
One can consider the average spiral as a self-consistent field created by all dipoles.
To find the  average pitch of the spiral  let us consider a single dipole 
with field $\delta{\bf n}$ given by (\ref{dn}) in a background field 
${\bf n}_b={\bf n}_0+\delta{\bf n}_b$ where
\vspace{-10pt}
\begin{equation}
\label{bg}
\delta{\bf n}_b=-\lambda{\overline m}_b({\bf e}_b\cdot{\bf r}) \ .
\end{equation}
Here $\delta{\bf n}_b$ is the self-consistent field of the dipoles,
${\bf e}_b=(1/\sqrt{2},\pm 1/\sqrt{2})$ is a unit vector orthogonal to the face 
of MBZ, and $\lambda$ is a parameter. The  interaction of the dipole with the background 
field is given by:
$\frac{\rho_s}{2}\int (\nabla\delta{\bf n}+\nabla\delta{\bf n_b})^2d^2r
-\frac{\rho_s}{2}\int (\nabla\delta{\bf n})^2d^2r
-\frac{\rho_s}{2}\int (\nabla\delta{\bf n}_b)^2d^2r$, which
 simply amounts to
$\rho_s\int (\nabla\delta{\bf n})(\nabla\delta{\bf n}_b)d^2r 
=-M\rho_s({\overline{\bf m}}\cdot{\overline{\bf m}}_b)({\bf e}\cdot {\bf e}_b)$.
Clearly the interaction energy has a minimum at ${\overline{\bf m}}={\overline{\bf m}}_b$ 
and ${\bf e}= {\bf e}_b$. The total energy at a finite concentration $x$ 
consists of the energy of each particular impurity (\ref{ev1}),
the interaction energy, and  the elastic energy of the background:
\vspace{-0pt}
\begin{equation}
\label{et}
E_{\lambda}=Ex -\rho_s C M \lambda x +\frac{\rho_s}{2}\lambda^2 \ .
\end{equation}
In the  interaction energy term we have introduced the  finite-size correction constant $C$.
Indeed,  Eq.~(\ref{dn}) is valid only at  very large distances from the impurity.
However  at a  finite distance the effective dipole moment is reduced,
according to Eq.~(\ref{solthet}), by the amount  $C=\left[1-e^{-2\kappa r}(1+2\kappa r)\right]$.
Substituting $r=1/\sqrt{\pi x}$ and $\kappa\approx0.4$, we find  for $x=0.03-0.05$
the value  $C\approx 0.7$.
Minimizing (\ref{et}) with respect to $\lambda$ we find $\lambda=CMx$.
Hence the average pitch is
\begin{equation}
\label{dq2}
\delta {\bf Q}=\lambda {\bf e}_b=C\frac{Zt}{\rho_s}\ x \ (1,\pm 1) \ .
\end{equation}
This expression determines the incommensurate shift of the neutron peak and  agrees  
well with the experimental data (\ref{data}) since $CZt/\rho_s \approx 4.1$.

The last question we want to discuss is the microscopic origin of the correlation  
length $l_{M}$.
First we notice that without randomness we would
 automatically have $l_{M}=\infty$, whereas experimentally  this quantity
 is about $l_{M} \approx 25-40\textrm{\AA}$ \cite{Fujita}.
 It has been suggested in Ref.~\cite{HCS} that topological defects related
to the random positions of impurities can destroy the long-range spiral order. 
This is a possible scenario, however we suggest a different mechanism.
In our opinion even randomly distributed dipoles (\ref{dn}) would create a true
long-range spiral order, similarly to a 2D ferroelectric.
 However,  our main observation is that the situation 
is {\it not} fully described by the point-like dipoles. Each impurity has a 
{\it finite size core} (see Eq.~(\ref{solthet}))
with diameter  $1/\kappa \sim 3-5$ lattice spacings
(depending on doping).
Therefore, given a random distribution of positions, there is always a finite probability
of impurity overlap. As soon as the impurities  overlap, a two-hole
``molecule'' is formed and the situation  changes dramatically. In the ``molecule'' the
Pauli blocking starts to play a role and in order to minimize the  energy the holes prefer to occupy
different pockets in  momentum space. If two pockets are occupied the (1,0) spiral
has lower energy, and  this is exactly what happen in the conducting phase \cite{SK}.
Hence such a ``molecule'' has a local spiral along (1,0) or (0,1) direction.
This spiral frustrates the (1,1) background and there is always a finite concentration
of such frustrating dipoles. Hence the ``molecule'' dipoles destroy the (1,1)
background similarly to the way  the ``atomic'' (single) dipoles destroy the N\'eel background at $x<0.02$.
One can consider these ``molecules'' as a precursor to the transition to the conductor
where the (1,0) spiral is realized. According to this picture the spin-glass
correlation length $l_{M}$ is large (but always finite) at very small $x$ and it should decrease
dramatically towards the percolation point $x\approx 0.055$ where the ``molecular''
configurations are becoming more important. This is exactly what is observed in experiment,
as seen in Fig.~6 of Ref.~\cite{Fujita}. In the superconducting phase the magnetic correlation length
should increase very rapidly, since theoretically it is infinity 
in  the fully uniform, metallic phase \cite{SK}. Indeed, experimentally
the correlation length quickly approaches the uniform limit \cite{Fujita} (it is 
  $>200 \textrm{\AA}, \ x=0.12$).

In conclusion, we have developed a description of the magnetic properties of underdoped
La$_{2-x}$Sr$_x$CuO$_4$, based on the extended $t-J$ model.
The theory describes the incommensurate elastic neutron scattering
above and below the metal-insulator transition at $x=0.055$. In particular it explains
why the incommensurate peak position rotates by 45$^\circ$  exactly at the insulator-metal transition.
The theory does not contain any fitting parameters, and the positions of the neutron peaks
both in the conducting (\ref{qc}) and in the insulating  (\ref{dq2}) phases, as well
as the critical concentration for destruction of the N\'eel order,  
follow from the calculated parameters of the extended $t-J$ model.
We also note that in La$_{2-x}$Sr$_x$CuO$_4$ static charge modulation (stripes) has 
 not been directly observed, suggesting that it is very weak  or not  present at all. 
We thus  believe that  a theory based on spiral magnetic
 correlations and no charge order
is fully sufficient to  describe the phenomena in this material.

We are grateful to  J. Haase, A. H. Castro Neto, G. V. M. Williams, 
O. K. Andersen, M. Ain, P. A. Lee, and Y. S. Lee  for important discussions
and comments. The support of the  Swiss National Fund is acknowledged (VNK).

\end{document}